\documentclass[aps, pre, floatfix, twocolumn, superscriptaddress, showpacs]{revtex4}
\usepackage{calc}
\usepackage{array}
\usepackage{graphicx}
\usepackage{amsmath}
\usepackage{hyperref}
\def\JPA{J. Phys. A: Math. Gen.}
\DeclareMathOperator{\Tr}{Tr}
\begin{document}

\title{Quantum mechanical potentials related to the prime numbers and Riemann zeros}

\author{D{\'{a}}niel \surname{Schumayer}} \email{dschumayer@physics.otago.ac.nz}
\affiliation{Jack Dodd Centre for Photonics and Ultra-Cold Atoms,
             Department of Physics, \\
             University of Otago,
             730 Cumberland St,
             Dunedin 9016,
             New Zealand}

\author{Brandon P. \surname{van Zyl}}
\affiliation{Department of Physics,
             St. Francis Xavier University, \\
             Antigonish, Nova Scotia B2G 2W5,
             Canada}

\author{David A. W. \surname{Hutchinson}}
\affiliation{Jack Dodd Centre for Photonics and Ultra-Cold Atoms,
             Department of Physics, \\
             University of Otago,
             730 Cumberland St,
             Dunedin 9016,
             New Zealand}
\date{\today}

\begin{abstract}
   Prime numbers are the building blocks of our arithmetic, however,
   their distribution still poses fundamental questions. Bernhard
   Riemann showed that the distribution of primes could be given
   explicitly if one knew the distribution of the non-trivial zeros of
   the Riemann $\zeta(s)$ function. According to the Hilbert-P{\'o}lya
   conjecture there exists a Hermitean operator of which the
   eigenvalues coincide with the real part of the non-trivial zeros of
   $\zeta(s)$. This idea encourages physicists to examine the
   properties of such possible operators, and they have found
   interesting connections between the distribution of zeros and the
   distribution of energy eigenvalues of quantum systems. We apply the
   Mar{\v{c}}henko approach to construct potentials with energy
   eigenvalues equal to the prime numbers and to the zeros of the
   $\zeta(s)$ function. We demonstrate the multifractal nature
   of these potentials by measuring the R{\'e}nyi dimension of their
   graphs. Our results offer hope for further analytical progress.
\end{abstract}
\maketitle
\section{Primes, zeros and quanta}
\label{sec:introduction}

The prime numbers are the building blocks for the positive integers,
since the fundamental theorem of arithmetic states that every positive
integer can be written as a product of primes, and this product is
unique up to a rearrangement of the factors. Additionally, not only
does the product of prime numbers have this remarkable property, but
also their sum. Only a decade ago O. Ramar{\'e} proved
\cite{Ramare1995} that any positive integer can be written as a sum of
no more than six prime numbers, but the Goldbach conjecture
\cite{Goldbach1742}, whether every number is expressible as a sum of
two prime numbers, remains unproven.

Based on empirical evidence, many mathematicians conjectured that the
prime counting function, $\pi(x)=\vert \lbrace p \,\vert\, p
\hspace*{1mm} {\mbox{is prime and}}\hspace*{1mm} p \le x \rbrace
\vert$, asymptotically behaves as the logarithmic integral
$\mathrm{Li}(x)$. Hadamard \cite{Hadamard1896} and de la Vall{\'e}e
Poussin \cite{Poussin1896} independently gave the rigorous proof for
this statement. Riemann derived the following exact formula
\cite{Schroeder1984}
\begin{equation} \label{eq:RiemannExpressionOfPi}
   \pi(x) = \lim_{x \rightarrow \infty}%
                 {\!\left ( R(x) - \sum_{\rho}{R(x^{\rho})} \right )}
\end{equation}
where $R(x)$ is the so-called Riemann function defined as $R(x) =
\sum_{m=1}^{\infty}{\mu(m)\mathrm{Li}(x^{1/m})}/m$, and $\mu(m)$
denotes the M{\"o}bius function. The sum in
(\ref{eq:RiemannExpressionOfPi}) is extended over all non-trivial
zeros, $\rho$, of the Riemann $\zeta(s)$ function, counted with their
multiplicities. The latter function $\zeta(s)$ is defined by the
infinite series, $\zeta (s) = \sum_{n=1}^{\infty}{n^{-s}}$ for $s>1$,
and, otherwise, by its analytic continuation over the complex $s$
plane.

Exploring the locations of the zeros of  $\zeta(s)$, Riemann made
his famous conjecture: all the non-trivial zeros lie on the $s =
\frac{1}{2} + {\mathrm{i}}t$ ($t$ is real) critical line. Proving or
disproving the Riemann-hypothesis remains the most tantalising
challenge in number theory since David Hilbert nominated it in 1900
\cite{Hilbert1900} as the eighth problem on his famous list of
compelling problems in mathematics
\footnote{See http://www.claymath.org/millennium/Riemann\_Hypothesis}.

The connection between the Riemann-hypothesis and physics seems to
date back to the early years of quantum mechanics. According to the
Hilbert-P{\'o}lya conjecture, the zeros of $\zeta(s)$ can be the
spectrum of an operator, ${\cal{O}} = \frac{1}{2}\, {\cal{I}} +
{\mathrm{i}} {\cal{H}}$, where ${\cal{H}}$ is self-adjoint. This
operator ${\cal{H}}$ might have a physical interpretation as a
Hamiltonian of a physical system and, therefore, the key to the proof
of the Riemann Hypothesis may have been coded in physics.

The analogy between the properties of $\zeta(s)$ and the energy
eigenvalues of a quantum mechanical system provide us some information
about the form of a possible operator ${\cal{H}}$ \cite{Rosu2003}. One
of these similarities, the comparison of the number of $\zeta(s)$
zeros and the number of energy eigenvalues below a threshold, suggests
that the physical system is quasi-one-dimensional. This link is
further strengthened by checking different statistics of the zeros,
such as the nearest-neighbour spacings, the $n$-correlations between
the zeros, etc. Montgomery showed that these distributions are all in
good agreement with the Gaussian unitary ensemble of random matrix
theory \cite{Montgomery1972}.

In this work we utilize an inverse scattering formalism and construct
potentials of which the energy eigenvalues are the zeros of the
Riemann $\zeta(s)$ function. We also consider the problem when the
eigenvalues are taken to be the prime numbers themselves. In Sec.
\ref{sec:InverseScatteringFormalism} we introduce our numerical method
and give evidence that it is capable of generating potentials from
sets of discrete energy eigenvalues, such as a finite set of $\zeta$
zeros or a finite set of prime numbers. We calculate the R{\'{e}}nyi
dimension \cite{Renyi1960} for these potentials. The results
anticipate that these potentials have multi-fractal nature. In section
\ref{sec:Comparison} we attempt to further clarify why previous
studies presented contradictory results for the fractal dimension.

\section{Inverse Scattering Formalism}
\label{sec:InverseScatteringFormalism}

Provided the Hilbert-P{\'o}lya conjecture is true, the natural and
plausible approach to finding operator ${\cal{H}}$ would be to
approximate it from a finite number of eigenvalues. We will follow
this path assuming the existence of a local potential ${\cal{V}}$
whose spectrum is related to the zeros of the Riemann $\zeta(s)$
function or, later, to the prime numbers.

The Mar{\v{c}}henko approach aims to reconstruct a symmetric
potential using the characteristics of both the bound states (energy
eigenvalues and normalisation constants) and the scattering states
(reflection coefficient at all energies). The question of the existence
and uniqueness of any solution obtained from the inversion procedures
is delicate, although if one assumes a one-dimensional, symmetric
potential the complete set of eigenvalues uniquely determines the
potential \cite{Barcilon1974}. Different, but mathematically equivalent,
methods exist \cite{Chadan1977} for reconstructing the scattering
potential in a one-dimensional quantum mechanical problem.

For a given set of energy eigenvalues and reflection coefficient the
quantum potential can be calculated from
\begin{equation}
   V(x) = - 2 \, \frac{\mathrm{d}}{\mathrm{d}x} K(x,x')\vert_{x'=x}
\end{equation}
where the order of operations is important. One should localise $x'$
first and then perform the differentiation. The function $K(x,x')$ is
the solution of the Mar{\v{c}}henko integral equation ($x'>x$)
\begin{equation} \label{eq:MarchenkoIntegralEquation}
   K(x,x') + K_{0}(x+x') 
           + \int_{x}^{\infty}{K(x,s)K_{0}(s+x')ds} 
           = 0
\end{equation}
and the kernel $K_{0}(z)$ is determined by the
spectral parameters
\begin{equation} \label{eq:MarchenkoKernelDefinition}
   K_{0}(z) = \frac{1}{2\pi} 
              \int_{-\infty}^{\infty}{R(k) e^{ikz}\,dz} 
              +
              \sum_{n=1}^{N}{c_{n} e^{-\kappa_{n} z}}.
\end{equation}
The input data are the reflection coefficient, $R(k)$, at energy $E =
\hbar^{2}k^{2}/2m$, and the normalisation constant, $c_{n}$, of the
$n$th bound state related to the discrete energy eigenvalue $E_{n} = -
\hbar^{2} \kappa_{n}^{2}/2m$.

In general, scattering states contribute to the kernel, and therefore
a family of potentials can be associated with a given set of energy
eigenvalues. However, according to our physical picture there are no
scattering states in our case. The potential is theoretically
infinitely deep, because the set of prime numbers and the set of
$\zeta(s)$ zeros are infinite sets with no upper bound. Henceforward
we take $R(k) \equiv 0$.

In the case of the reflectionless potential, the scattering states in
equation (\ref{eq:MarchenkoKernelDefinition}) do not contribute to
the kernel and, therefore, lead to a separable and exactly solvable
integral equation. The potential can then be obtained from the
following formula \cite{Kay1956}
\begin{subequations}
   \begin{equation} \label{eq:ISTGeneralExpressionForThePotential}
      V(x) = -2 \frac{{\mathrm{d}}^{2}}%
                     {{\mathrm{d}x^{2}}}
                \ln{ \!\left (
                              \det{( {\mathbf{I}} + {\mathbf{C}})}
                       \right )
                   }
   \end{equation}
where ${\mathbf{I}}$ denotes identity matrix, and
   \begin{eqnarray} 
      \label{eq:MatrixC}
      {\mathbf{C}}_{m,n} &=& \frac{c_{m} c_{n}}%
                                {\kappa_{m} + \kappa_{n}} \,
                           e^{-(\kappa_{m}+\kappa_{n}) x},
      \\
      \label{eq:NormalisationConstants}
      \frac{c_{n}^{2}}{2\kappa_{n}} &=& \prod_{\stackrel{\scriptstyle m = 1}{m \ne n}}^{N}%
                                      {
                                       \left \vert
                                       \frac{\kappa_{m} + \kappa_{n}}
                                       {\kappa_{m} - \kappa_{n}}
                                       \right \vert
                                      }
   \end{eqnarray}
\end{subequations}
Choosing a set of eigenvalues, $\left \lbrace \kappa_{n} \right
\rbrace$, one may calculate the corresponding normalisation constants,
$\left \lbrace c_{n} \right \rbrace$ and matrix ${\mathbf{C}}$ from
equations (\ref{eq:MatrixC}-c). Substituting this matrix into
(\ref{eq:ISTGeneralExpressionForThePotential}), the desired potential
can be calculated.

We note here briefly, that using the matrix identity,
$
   \ln{(\det{\left ( {\mathbf{M}} \right )})}
   =
   \Tr{ \left (
                              \ln{( {\mathbf{M}} )}
                      \right )
                    }
$,
and the power series expansion of $\ln{(1+x)}$,
one can symbolically derive the following expression
for the potential
\begin{equation} \label{eq:FormalPowerSeries}
   V(x) = 2 \sum_{r=1}^{\infty}%
                 {\frac{(-1)^{r}}{r}
                  \Tr{\left (
                         \frac{\mathrm{d}^{2}}%
                              {\mathrm{d}x^{2}}
                         \,
                         {\mathbf{C}}^{r}
                      \right )
                     }
                 }
\end{equation}
where $\Tr{(\mathbf{M})}$ denotes the trace of matrix $\mathbf{M}$.

The formulae (\ref{eq:ISTGeneralExpressionForThePotential}-c) above
form the basis of our calculations. Although these expressions may
seem simple, the accurate numerical evaluation of the determinant
can prove challenging, particularly as the number of eigenvalues
is increased.

The inversion technique in its present form is numerically not
convenient for more than about five hundred eigenvalues, for two
reasons: firstly the matrices involved are dense, and secondly, the
numerical precision required is demanding, since for medium
values of $x$ one has to calculate the exponential functions in
(\ref{eq:MatrixC}) very accurately to have precise cancellation.
The transformation of the formulae
(\ref{eq:ISTGeneralExpressionForThePotential}-c) into a numerically
more tractable form is under investigation.

For this reason, for large sets of eigenvalues, we used the
dressing-transformation \cite{Ramani1995} to calculate the potential.
We have checked numerically, up to three hundred energy eigenvalues,
that Mar{\v{c}}henko's inversion method and the dressing-transformation
give identical results within numerical accuracy. The strength of our
approach is in the explicit formulae for the potential construction.
Using the dressing-transformation one has to recursively solve
ordinary differential equations, since the potential is built up
by incorporating the energy eigenvalues one by one, so in every step
the solution of the previous step is used. Therefore, the 
applicability of this method to gain general and analytical results
is limited. Contrary, in our method all quantities are expressed in
terms of the input parameters, viz. the set of energy eigenvalues,
offering some hope of analytic progress.

\subsection{Reconstruction of well-known potentials}
\label{subsec:WellKnownPotentials}

To illustrate the method, we reconstruct well-known potentials from
their spectra, using the triangular and harmonic oscillator
potentials. 
\begin{figure*}[th!]
   \parbox{\textwidth}%
          {
           \includegraphics[angle=-90, width=\columnwidth]{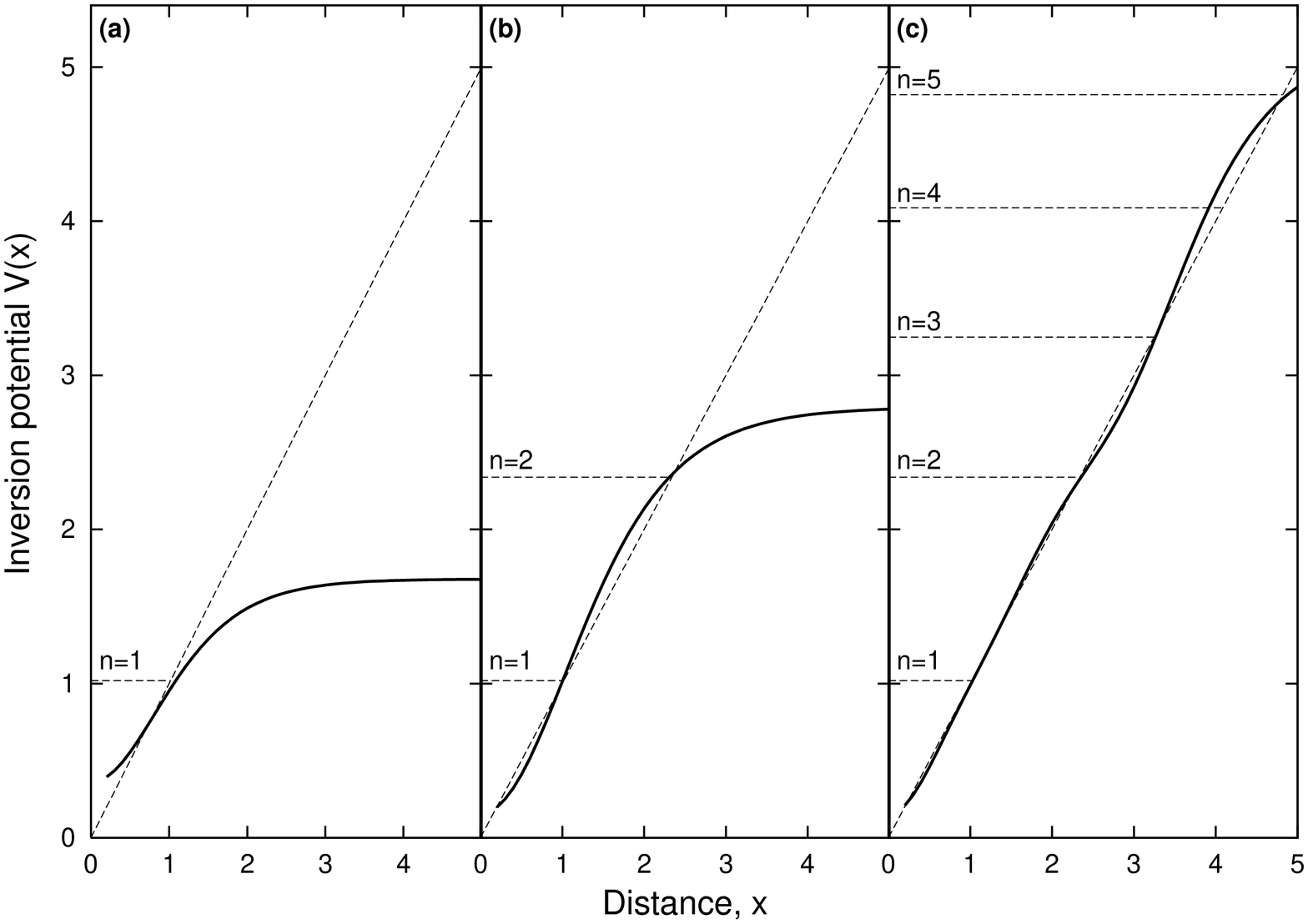}
           \includegraphics[angle=-90, width=\columnwidth]{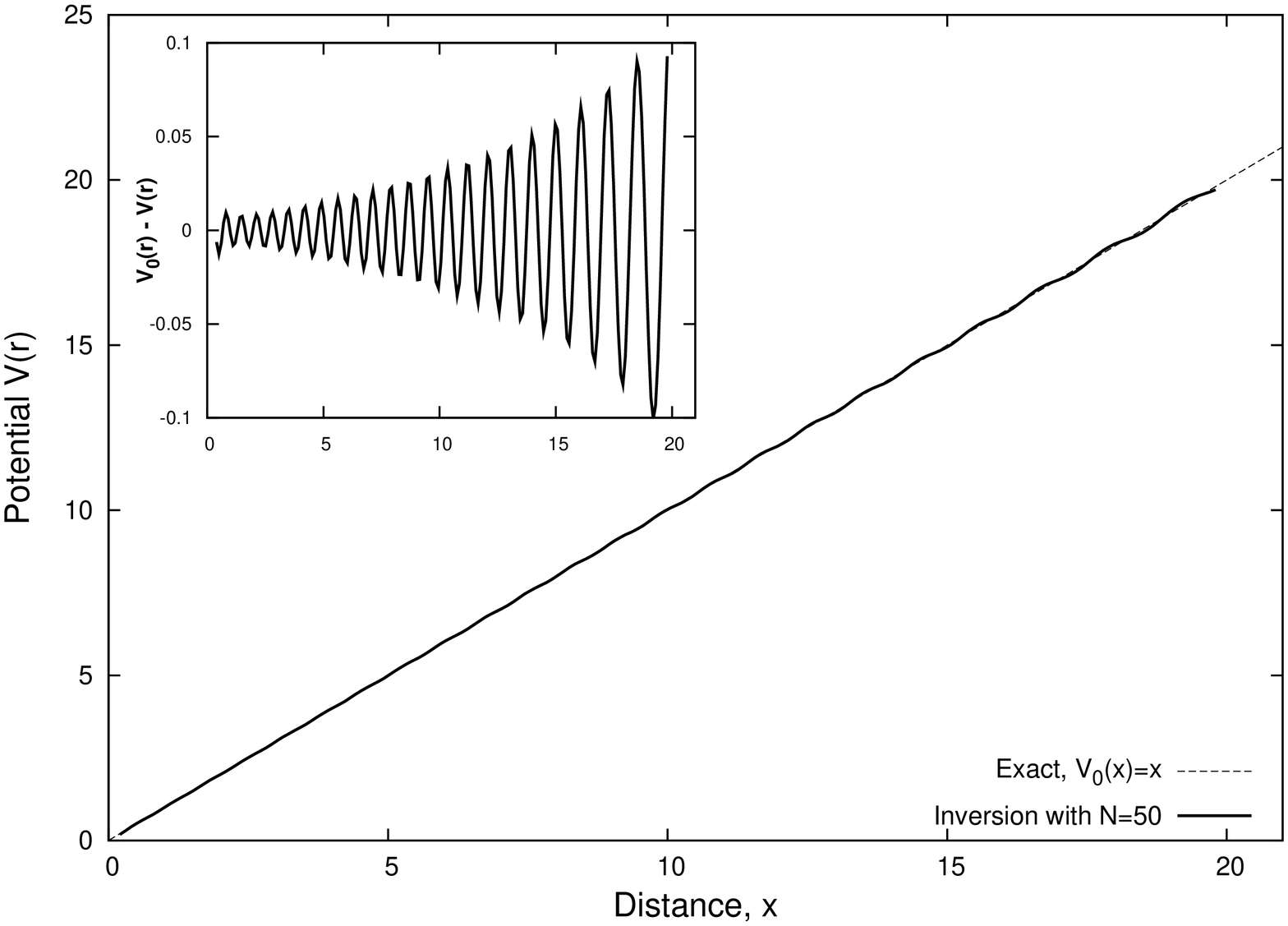}
           \caption{\label{fig:Reference_Linear}
                    The reference potential, $V_{0}(x) = x$
                    (dashed line), and the inversion potentials (solid
                    lines), $V(x)$, are shown using (a) one (b) two and
                    (c) five energy eigenvalues indicated with horizontal
                    dashed lines. The lower figure depicts the reference
                    and inversion potential derived using the first fifty
                    energy eigenvalues; the inset illustrates the difference
                    between $V_{0}(x)$ and $V(x)$.}
           }
   \parbox{\textwidth}%
          {
           \includegraphics[angle=-90, width=\columnwidth]{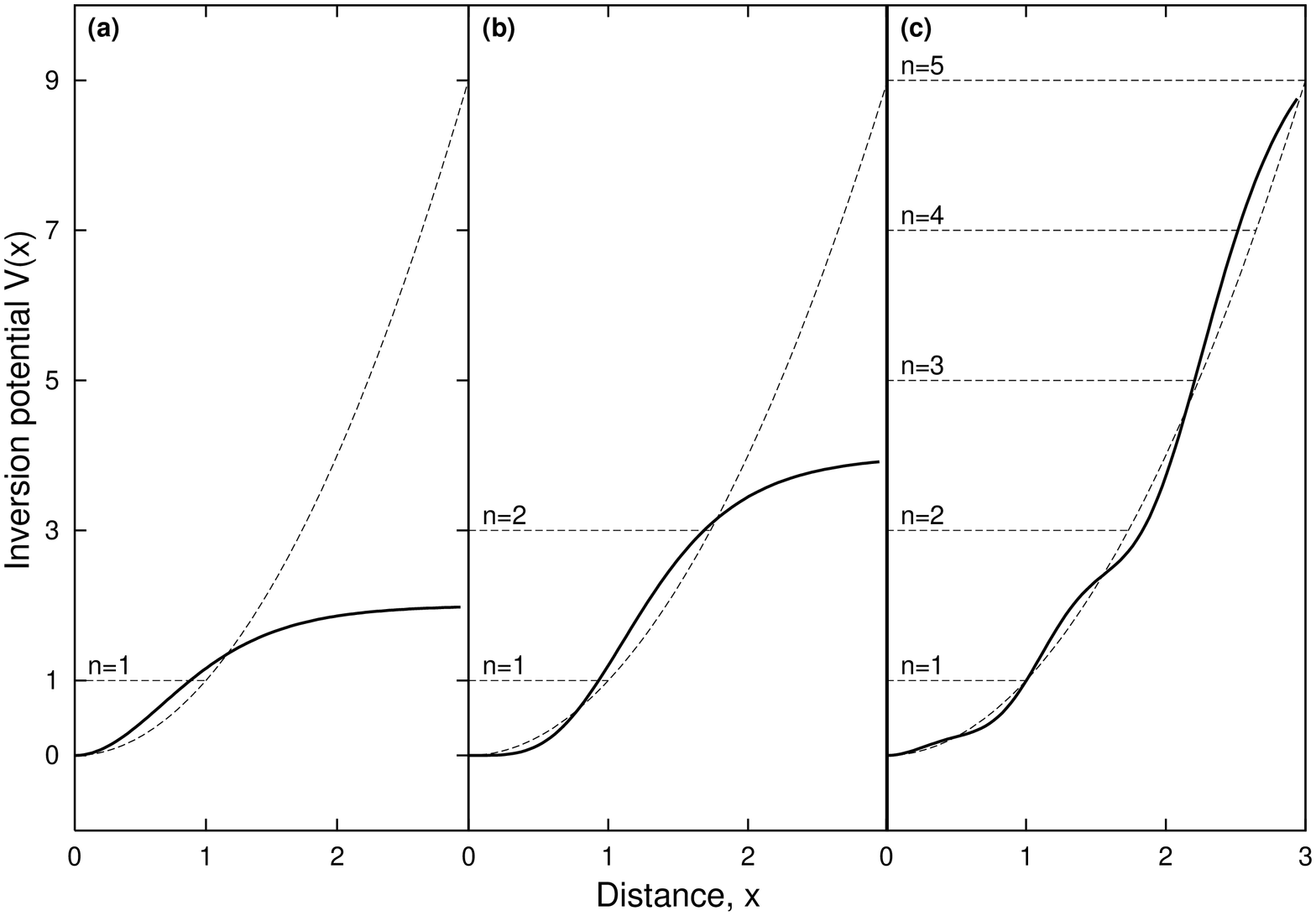}
           \includegraphics[angle=-90, width=\columnwidth]{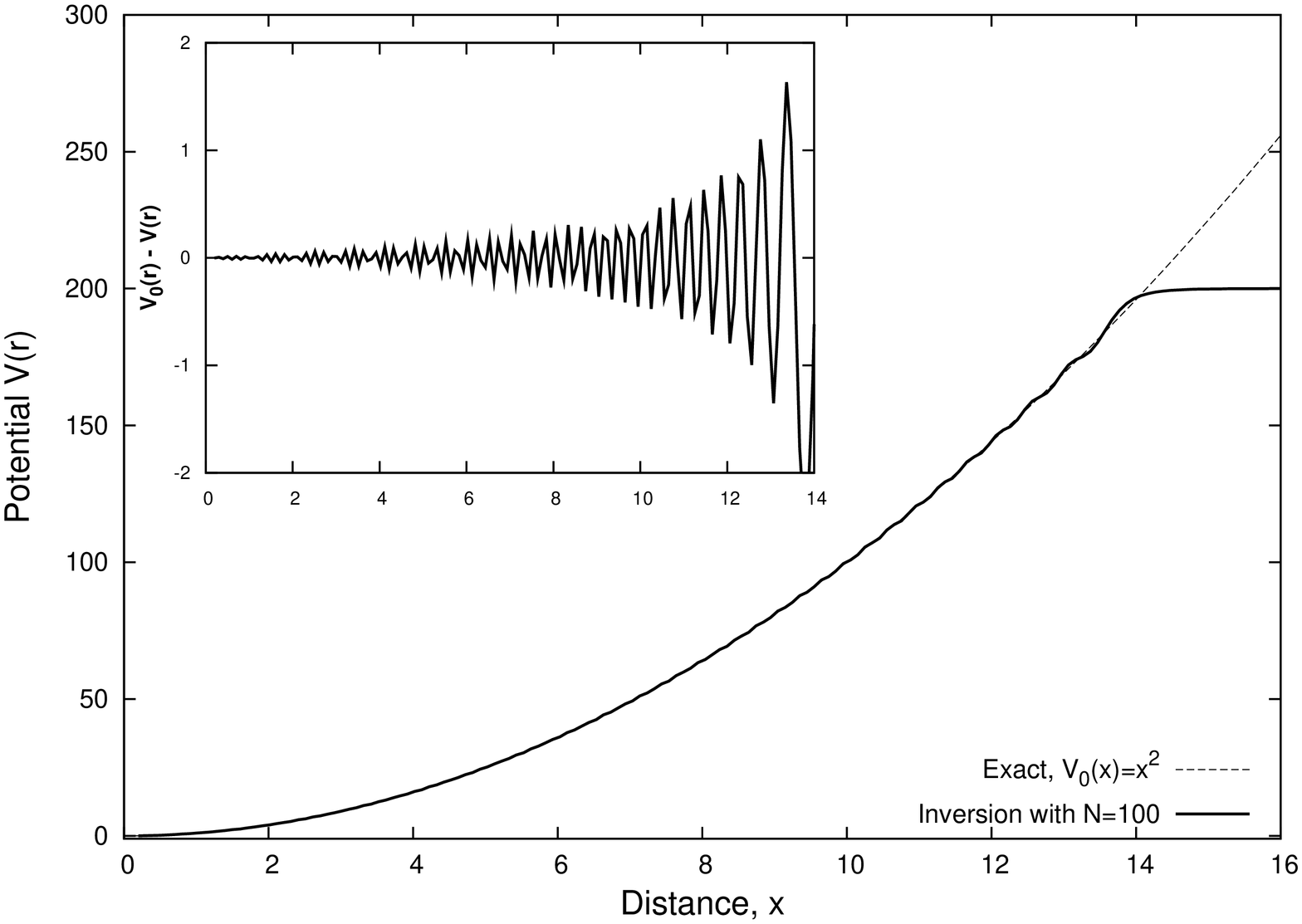}
           \caption{\label{fig:Reference_Harmonic}
                    Figures show the reference potential, $V_{0}(x) = x^{2}$,
                    (dashed line) and the inversion potential, $V(x)$, (solid
                    line) using (a) one (b) two and (b) five energy
                    eigenvalues which are indicated with horizontal dashed
                    lines. In the lower panel, the same quantities are
                    presented with the first one hundred energy eigenvalues
                    utilised. The inset depicts $V_{0}(x) - V(x)$.}
          }
\end{figure*}
Later the extended Numerov method \cite{Fack1987} is used to
calculate the energy eigenvalues of the inversion potentials.
This also serves to check the validity of the potentials obtained
using the Mar{\v{c}}henko approach.

\begin{table}[bht!]
   \caption{\label{table:ComparisonExactAndInversionPotentials}
            Comparison of the known energy eigenvalues,
            $\epsilon_{0, n}$, for the harmonic and triangular reference
            potentials with the energy eigenvalues, $\epsilon_{n}$,
            of the appropriate inversion potentials calculated for
            the first one hundred eigenvalues. The table shows the
            first and the last five eigenvalues.}
   \begin{tabular}{r<{\hspace*{2mm}}rrp{5mm}rr}
      & \multicolumn{2}{c}{
           \parbox[b]{30mm}{\centering Harmonic potential}} & &
        \multicolumn{2}{c}{
           \parbox[b]{29mm}{\centering Triangular potential}}  \\
      \cline{2-3} \cline{5-6}
      n & $\epsilon_{0,n}$ & $\epsilon_{n}$ & & 
          $\epsilon_{0,n}$ & $\epsilon_{n}$ \\
      \hline
        1  &   1   &   1.001923 & &  1.018793 &  1.015439  \\
        2  &   3   &   3.000020 & &  2.338107 &  2.338100  \\
        3  &   5   &   5.000944 & &  3.248198 &  3.247152  \\
        4  &   7   &   7.000031 & &  4.087949 &  4.087942  \\
        5  &   9   &   9.000696 & &  4.820099 &  4.819397  \\
           &       &            & &           &            \\
       96  & 191   & 190.997520 & & 36.995074 & 36.995066  \\
       97  & 193   & 192.996641 & & 37.252699 & 37.252615  \\
       98  & 195   & 194.996021 & & 37.509795 & 37.509785  \\
       99  & 197   & 197.015433 & & 37.765659 & 37.765580  \\
      100  & 199   & 198.984293 & & 38.021009 & 38.021020  \\
      \hline\hline
   \end{tabular}
\end{table}

We build up the potentials using a finite number of eigenvalues,
starting with one then two, five and finally one hundred eigenvalues
from the bottom of the known spectrum. As one may expect,
incorporating more and more eigenvalues into the method results in
the inversion potential becoming more and more accurate and 
reproducing the spectrum faithfully. This tendency is clearly
captured in Figure (\ref{fig:Reference_Linear}) for the triangular
potential and in Figure (\ref{fig:Reference_Harmonic}) for the
harmonic potential. Furthermore, the inversion potentials reach
their asymptotic value exponentially \cite{Kay1956} and this
asymptote lies between the last energy eigenvalue used for the
inversion and the next eigenvalue. Although the discrepancy
between the exact and inversion potentials becomes larger toward
the edge of the inversion potential, the energy eigenvalues are
still correctly reproduced (see Table
\ref{table:ComparisonExactAndInversionPotentials}) with tolerable
errors.

\subsection{Inversion potential for prime numbers}
\label{subsec:PotentialForPrimeNumbers}

Using semi-classical arguments, one may show that for a
one-dimensional potential the energy eigenvalues cannot increase
more rapidly than quadratically, i.e. $\epsilon_{n} \sim n^{2}$.

Intuitively this may be seen by noting that in the case of the
triangular attractive potential $\epsilon_{n}$ scales as $n^{2/3}$,
while for the harmonic oscillator $\epsilon_{n}$ goes as $n$, and,
as a limiting case, for the infinite-box potential, $\epsilon_{n}$
varies as $n^{2}$. A corollary of the Hadamard-Poussin theorem
\cite{Titchmarsh1986} is that the $n$th prime number is approximately
$n\ln{\!(n)}$, which is clearly less than $n^{2}$. We cannot,
therefore, rule out the existence of a quantum mechanical potential
which has prime numbers as energy eigenvalues.

We now turn to the construction of a semi-classical potential for
which the first $n$ energy eigenvalues coincide with the first $n$
prime numbers. There is no theoretical limit on the number of
incorporated prime numbers, although numerically the calculation
becomes quite cumbersome.

\begin{table}
    \caption{\label{table:ComparisonExactAndInversionPotentials2}
             The energy eigenvalues of the potentials derived for the
             prime numbers and the zeros of the $\zeta(s)$ function.
             For both the primes and zeros of the Riemann $\zeta(s)$
             function, the first column comprises the exact
             eigenvalues, $\epsilon_{0,n}$, the second one contains
             $\epsilon_{\mathrm{sc},n}$ for the semi-classical
             potential, and the last incorporates the energy
             eigenvalues $\epsilon_{n}$ of the appropriate inversion
             potential.}
   \begin{tabular}{rrrrcrrr}
           & \multicolumn{3}{c}{Prime numbers} & & \multicolumn{3}{c}{Riemann $\zeta(s)$ zeros} \\
      \cline{2-4} \cline{6-8}
        n  & $\epsilon_{0,n}$ & $\epsilon_{{\mathrm{sc}}, n}$ & $\epsilon_{n}$ & & $\epsilon_{0,n}$ & $\epsilon_{{\mathrm{sc}}, n}$ & $\epsilon_{n}$\\
      \hline
        1  &   2  &   0.6895 &  1.6387  &  &   14.1347 &  13.4690 & 13.0302   \\
        2  &   3  &   2.5316 &  3.0005  &  &   21.0220 &  23.2274 & 21.0208   \\
        3  &   5  &   5.0674 &  4.7052  &  &   25.0109 &  29.8790 & 24.7026   \\
        4  &   7  &   7.9717 &  7.0006  &  &   30.4249 &  36.0644 & 30.4234   \\
        5  &  11  &  11.1201 &  10.702  &  &   32.9351 &  41.4187 & 32.8091   \\
                  &          &          &  &           &          &           \\
       96  & 503  & 513.8440 & 503.0008 &  &  229.3374 & 284.3914 & 229.3354  \\
       97  & 509  & 520.3027 & 508.7052 &  &  231.2502 & 287.1530 & 231.2259  \\
       98  & 521  & 526.7728 & 520.9981 &  &  231.9872 & 289.9657 & 231.9865  \\
       99  & 523  & 533.2544 & 522.9371 &  &  233.6934 & 292.8088 & 233.6322  \\
      100  & 541  & 539.7472 & 540.9843 &  &  236.5242 & 295.7020 & 236.5215  \\
      \hline\hline
   \end{tabular}
\end{table}

Using the Wentzel-Kramers-Brillouin semi-classical quantisation
formula \cite{Galindo1991} and the leading terms of the prime number
counting function, $\pi(E) \approx R(E)$, one may derive 
\cite{Mussardo1997} the following implicit equation for a potential
of which the eigenvalues are approximately the prime numbers ($E_{0}
\ge 1$):
\begin{equation} \label{eq:SemiclassicalForPrimeNumbers}
   x(V) = \sum_{m=1}^{\infty}{
                             \frac{\mu(m)}{m}
                             \int_{E_{0}}^{V}{
                                             \frac{E^{\frac{1-m}{m}}}%
                                                  {\ln{\!(E)}
                                                   \sqrt{V-E}
                                                  }
                                             \,dE
                                             }
                             }
\end{equation}
where $E_{0}$ denotes the reference energy-level. Due to the density
of prime numbers, $\rho(E) \approx \ln{(E)}$, this reference energy
cannot be less than 1. Even though the integral, for general $m$,
cannot be expressed using elementary functions, one may bound the
integral from below and from above such that ($x \gg 1$)
\begin{equation}
  x^{2} \ln^{2}{(x)} < V(x)-E_{0} < x^{2} \ln^{2}{(x \ln(x))}.
\end{equation}

\begin{figure}
   \includegraphics[angle=-90, width=\textwidth/2]{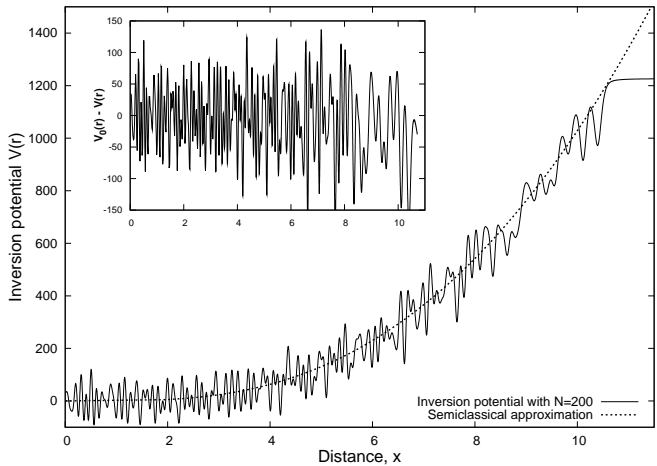}
   \caption{\label{fig:PrimePotential}
            The main figure depicts the semi-classical potential
            (dashed line), $V_{0}(x)$, and the inversion potential
            (solid line), $V(x)$, derived from using
            Mar{\v{c}}henko's method with the first two hundred
            prime numbers. The inset shows the difference of
            $V_{0}(x)$ and $V(x)$.}
\end{figure}

In Figure \ref{fig:PrimePotential} we plot the inversion potential
calculated from the first two hundred prime numbers and the
associated semi-classical potential from equation
(\ref{eq:SemiclassicalForPrimeNumbers}). It is apparent that the
inversion potential oscillates around the semi-classical potential
except close to the edge of the potential. Similarly, one may solve
the Schr{\"{o}}dinger equation with the semi-classical potential and
with the inversion potential obtained above, comparing how well they
reconstruct the original set of eigenvalues. Table
\ref{table:ComparisonExactAndInversionPotentials2} comprises a
selection from the original set of eigenvalues
($\epsilon_{\mathrm{0}, n}$), labeled by $n$, and the energy
eigenvalues of the semi-classical ($\epsilon_{\mathrm{sc}, n}$) and
inversion potential ($\epsilon_{n}$) with the same quantum number.
This also served as a numerical check of our method. The
semi-classical energy eigenvalues capture the trend, but -- as
expected -- those derived from the inversion potentials are much
better approximations to the exact eigenvalues. One may notice that
the agreement between $\epsilon_{\mathrm{0}, n}$ and $\epsilon_{n}$
is consistently much better for even values of $n$. The same tendency
can also be seen in Table
\ref{table:ComparisonExactAndInversionPotentials} for the triangular
and harmonic oscillator potential, although on an order of magnitude
smaller scale. The underlying reason for this effect is the subject
of ongoing investigation.

\subsection{Inversion potential for $\zeta(s)$ zeros}
\label{subsec:PotentialForZetaZeros}

Similarly to the semi-classical approximation derived for the prime
numbers, one may calculate a potential corresponding to the Riemann
$\zeta(s)$ zeros using the fact that the number of zeros
\cite{Titchmarsh1986}
\begin{equation} \label{eq:NumberOfZetaZeros}
   {\mathcal{N}} (E)
   =
   \frac{1}{2\pi} E \ln{\!(E)} -
   \frac{1+\ln{\!(2\pi)}}{2\pi} E +
   \frac{7}{8} +
   {\cal{O}} \!\left ( \ln{\!(E)} \right )
\end{equation}
Calculating the average density of the $\zeta(s)$ zeros from the
expression above restricts the choice of the otherwise arbitrary
reference energy level to $E_{0} \ge 2\pi$. Inserting the density
into the Wentzel-Kramers-Brillouin semi-classical quantisation
formula, we obtain (see 2.727.5 in \cite{Gradshteyn2000})
\begin{eqnarray} \label{eq:WuSprungPotential}
   x(V) =
          \frac{1}{\pi}
          \biggl \lbrack
          \sqrt{V-E_{0}}
          \ln{ \!\left ( \frac{E_{0}}{2\pi e^{2}} \right )}
          + \hspace*{15mm} \nonumber \\
          \label{eq:SemiclassicalForZetaZeros}
          \sqrt{V}
          \ln{\!\left ( \frac{\sqrt{V} + \sqrt{V-E_{0}}}%
                             {\sqrt{V} - \sqrt{V-E_{0}}}
                \right )
             }
          \biggr \rbrack,
\end{eqnarray}
which is identical to that given by Wu and Sprung \cite{Wu1993}. The
structure of the semi-classical potential close to the origin depends
on the choice of the reference energy level. If $E_{0} > 2\pi$ then
$V(x)-E_{0} \sim x^{2}$, but in case of $E_{0} = 2\pi$ the potential
grows as $V(x)-E_{0} \sim x^{2/3}$. As $\vert x \vert $ approaches
infinity the potential becomes independent of $E_{0}$ and expression
(\ref{eq:WuSprungPotential}) can be inverted to obtain the
following asymptotic
\begin{equation} \label{eq:WuSprungPotentialAsymptote}
   V(x)
   \approx
   \frac{\pi^2 x^2}{4}
   \left \lbrack
       {\mathrm{W}}
       \!\left (
                \sqrt{\frac{\pi}{2}}
                \frac{\vert x \vert}%
                     {e}
       \right )
   \right \rbrack^{-2}
\end{equation}
where $W(z)$ denotes the Lambert-W function.

\begin{figure}
   \includegraphics[angle=-90, width=\textwidth/2]{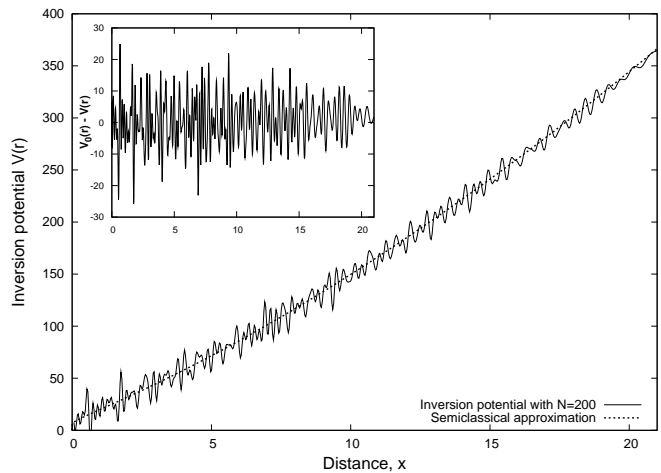}
   \caption{\label{fig:RiemannZetaPotential}
            Main figure shows the semi-classical potential (dashed
            line), $V_{0}(x)$, and the inversion potential (solid
            line), $V(x)$, derived from the inverse scattering method
            using the first two hundred energy eigenvalues. The inset
            depicts the difference of $V_{0}(x)$ and $V(x)$.}
\end{figure}

Applying the formulae (\ref{eq:ISTGeneralExpressionForThePotential}-c)
we calculated a number of potentials supporting the first $n$ zeros
of the $\zeta(s)$ as energy eigenvalues, utilising a tabulated form
of the low-lying zeros \cite{OdlyzkoWeb}. As an example, we have
plotted the potential for $N=200$ in Figure \ref{fig:RiemannZetaPotential}
and in Table \ref{table:ComparisonExactAndInversionPotentials2}
one can compare how well the energy eigenvalues of the inversion
potential coincide with the original eigenvalues, i.e. the zeros
of the Riemann $\zeta(s)$ function. In this case the agreement is
even better than it was for the prime numbers. This can be explained
by the much slower increase of the potential than that for prime
numbers, as $x$ approaches infinity. Similar effects are seen for
the two pedagogical examples in Table 
\ref{table:ComparisonExactAndInversionPotentials}. The energy
eigenvalues for the triangular potential are, at least, an order of
magnitude more accurate than those for the harmonic potential.

\section{Comparison with earlier results}
\label{sec:Comparison}

Both variational and dressing-transformation techniques, have
already been applied to construct quantum mechanical potentials
for which the energy eigenvalues are either the zeros of the Riemann
$\zeta(s)$ function \cite{Wu1993, Ramani1995, Zyl2003}, or the
prime numbers \cite{Zyl2003}. The common feature of these methods
is that the potential is built up in recursion by incorporating more
and more eigenvalues into the spectrum.

\begin{figure}
   \includegraphics[angle=-90, width=\textwidth/2]{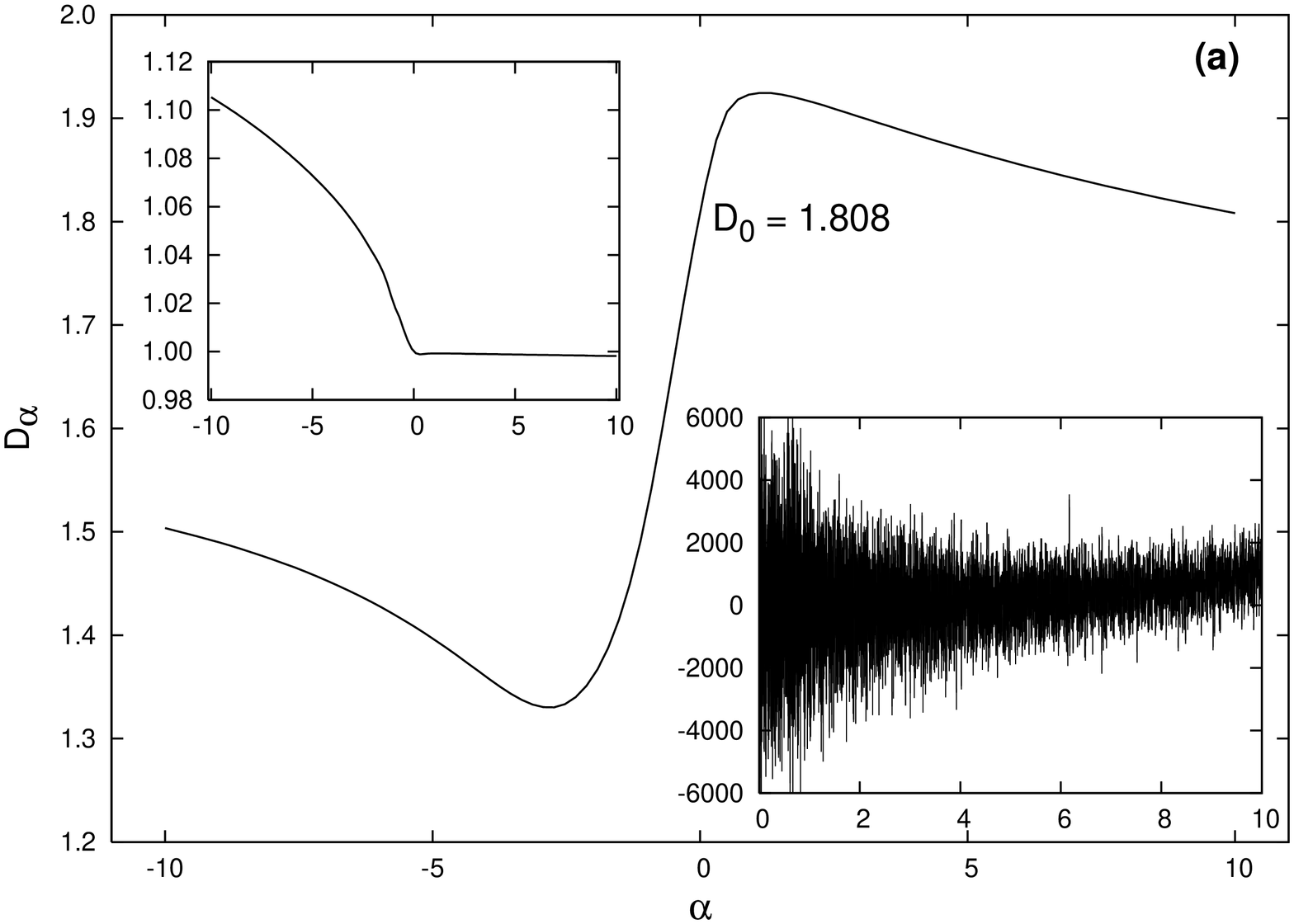}
   \includegraphics[angle=-90, width=\textwidth/2]{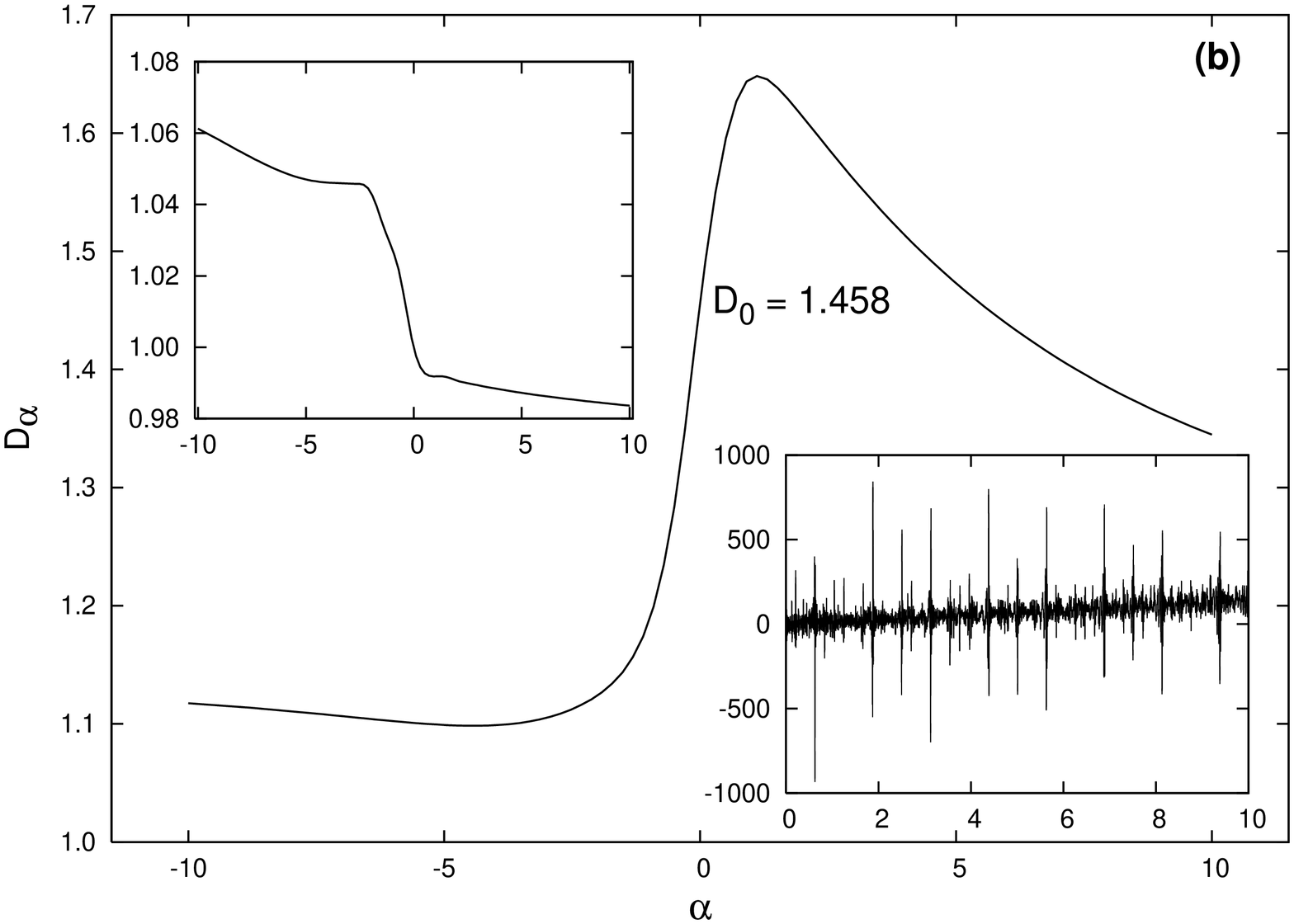}
   \caption{\label{fig:FractalDimension}
            The R{\'{e}}nyi-dimension, $D_{\alpha}$, is shown for the
            potentials belonging to (a) the prime numbers and (b) the
            Riemann $\zeta(s)$ zeros. Insets on the left show 
            $D_{\alpha}$ for the marginal distribution (see text)
            while those on the right depict the de-trended data.
            Here, in the dressing-transformation, we used the first
            $10^{5}$ eigenvalues and spatial step-size $h=10^{-6}$
            for both the prime numbers and the zeros of $\zeta(s)$.
            }
\end{figure}

Previous works \cite{Wu1993, Zyl2003} estimated the box-counting
dimension of the potentials belonging to the prime numbers to be
1.8, and for the Riemann $\zeta(s)$ zeros to be 1.5, where the
number of eigenvalues used ranged from 100 to 32000. Our
measurements broadly support these values (see $D_{0}$ in Figure
\ref{fig:FractalDimension}).

In order to reproduce these findings we treat the graph of
the potential as a signal. To measuring the fractal
dimension we de-trend the signal, i.e. subtract the actual
inversion potential from the semi-classical potential, $\xi(x)
= V(x) - V_{\mathrm{sc}}(x)$. Moreover, we limit ourselves to
the spatial range of $[0, 10]$ to eliminate any boundary effect
arising from the fact that both Mar{\v{c}}henko's method and the
dressing-transformation produce a potential with a constant 
asymptotic value for large spatial coordinates.

We have measured the R{\'{e}}nyi-dimension of the potentials
\cite{Renyi1960, Kruger1996}, defined as
\begin{equation} \label{eq:RenyiDimension}
   D_{\alpha}(X)
   =
   \frac{1}{\alpha - 1} 
   \lim_{\epsilon \rightarrow 0^{+}}%
       {\left [ \frac{\ln{\!\left ( \sum_{i=1}^{N}{p_{i}^{\alpha}} \right )}}%
             {\ln{(\epsilon)} }
        \right ]
       }
\end{equation}
where $p_{i}$ is the probability that the discrete random variable
$X$ falls into a box centered at $x_{i}$ with side $\epsilon$. This
probability can be approximated using the relative frequencies
obtained by dividing the embedding two-dimensional $(x,V)$ space into
a finite number of bins and counted how many times the potential
takes its value in the given box. Contrary to the general box-counting
method, which treats every box equally regardless of how many points
of the fractal a given box contains, if $\alpha > 0$ in
(\ref{eq:RenyiDimension}) then boxes with higher relative frequencies
will dominate the summand, therefore determining $D_{\alpha}(X)$. On
the other hand, if $\alpha <0$ then the formula weighs the less dense
boxes more and measures their scaling properties. As a special case,
$\alpha=0$ associates equal weights to every box, and therefore
$D_{0}$ should reproduce the box-counting dimension. It can also be
shown \cite{Renyi1960} that for $\alpha \rightarrow 1$ the numerator
including the pre-factor, converges to the Shannon-entropy, defined
to be $- \sum_{i}{p_{i} \ln{(p_{i})}}$. Concluding, by calculating
the generalised R{\'{e}}nyi dimension one can ``scan'' the fractal
by its ``density'', and measure its heterogeneity. If $D_{\alpha}$
varies with $\alpha$ then the fractal is a multi-fractal, since its
subsets scale differently.

One may derive two statistics for $\xi$ based on (\ref{eq:RenyiDimension});
(a) using a two-dimensional grid and approximating the
two-dimensional conditional probability distribution with the
relative frequencies of the individual boxes, or
(b) calculating the generalised dimension for the marginal
probability distribution of $\xi$.

In Figure \ref{fig:FractalDimension} both set of statistics
are demonstrated showing the generalised dimension for the
two-dimensional probability distribution and the insets depicting
$D_{\alpha}$ for the marginal distribution. The box-counting
dimensions, $D_{0}$, are also indicated on the graphs. Although
both curves in the main figures have an overall ``S'' shape, their
structure is different. In the case of the potential generated from
the prime numbers, both the positive and negative $\alpha$ branches
of the curve show monotonic decrease towards the right. However, for
the potential designed from the zeros of the Riemann $\zeta(s)$
function, the negative $\alpha$ branch of the curve remains nearly
constant.

These results suggest that the potentials calculated for the prime
numbers and for the zeros of the Riemann $\zeta(s)$ function are
indeed multi-fractals \cite{Wolf1989}. The steep middle part of the
curves also explains why earlier studies \cite{Wu1990, Wu1993,
Ramani1995, Zyl2003} differed in the box-dimension. The number of
incorporated energy eigenvalues strongly influence the conditional
probability associated to one box and eventually shift $D_{0}$.

Finally, we mention another conjectured property of the quantum
system supposed to possess the zeros of the Riemann $\zeta(s)$
function as energy eigenvalues, namely that ${\cal{H}}$ may violate
the time-reversal symmetry.

Similarly to the prime counting function, $\pi(x)$, one may define
a function, ${\mathcal{N}}(t)$, which counts the zeros of the Riemann
$\zeta(s)$ function, i.e. a function jumping by unity whenever $t$
passes over of the zeros, $t_{n}$. It is proven \cite{Titchmarsh1986}
that the function ${\mathcal{N}}$ can be decomposed into a smooth and
fluctuation part:
${\mathcal{N}}(t) = N(t) + {\mathcal{N}}_{\mathrm{osc}}(t)$,
where $N(t)$ has been given in (\ref{eq:NumberOfZetaZeros})
explicitly. The fluctuating term has remarkably similar structure to
Gutzwiller's trace formula \cite{Gutzwiller1971} giving the density
of states of a quantum system. The comparison of the two formulae led
to the hypothesis \cite{Berry1999} that a quantum system with the
zeros of the Riemann $\zeta(s)$ function as energy eigenvalues, does
not possess time-reversal symmetry. The approach presented in this
paper creates a symmetric, one-dimensional, although multi-fractal
potential, $V(x)$, for which the corresponding energy eigenvalues
coincide with the first $n$ non-trivial zeros of the Riemann
$\zeta(s)$ function. The reflection symmetry of the potential,
$V(x) = V(-x)$, guarantees time-reversal symmetry of the Hamiltonian,
${\mathcal{H}} = p^{2}/2m + V(x)$. This result, therefore, allows us
to assume the existence of a quantum system having the $\zeta$ zeros
as energy eigenvalue and obeying the time-reversal symmetry
simultaneously.

\section{Conclusion}
\label{sec:Conclusion}

In the present paper we used Mar{\v{c}}henko's method, one of the
inverse scattering methods, to construct one-dimensional, symmetric
quantum potentials, the energy eigenvalues of which coincide with
either the prime numbers or the zeros of the Riemann $\zeta(s)$
function. We have demonstrated the accuracy and usefulness of this
method on two pedagogical examples, the triangular and harmonic
potentials. For both cases we showed the reconstructed potentials and
calculated the energy eigenvalues, which agreed with the
predescribed values very well. Later, we applied the same technique
and numerically calculated potentials for the prime numbers and zeros
of $\zeta(s)$. We found that the outcome of the Mar{\v{c}}henko
method is identical to that of the dressing-transform used
previously. At the present stage, the latter method is numerically
preferable to the Mar{\v{c}}henko's method. Using the dressing-transform,
we created potentials, to high accuracy, from the first one-hundred
thousand prime numbers and also for the same number of zeros of the
Riemann $\zeta(s)$ function. Looking at the graphs of these potentials
as signals one can analyse their statistical properties. After de-trending
these signals we calculated the R{\'{e}}nyi-dimension, which is a
generalised fractal dimension. Our results suggest that inversion
potentials are multi-fractals for both the prime numbers and for the
zeros of $\zeta(s)$. The specific values of the generalised dimension
for the prime numbers $D_{0} = 1.808$, and for the $\zeta(s)$ zeros,
$D_{0} = 1.458$, agree well with \cite{Zyl2003}.

Even though Mar{\v{c}}henko's method is not yet able to compete
with the dressing-transform in the number of eigenvalues incorporated
into the potential, it gives explicit formulae for how one can build
up such potentials without recursion. Looking at formula
(\ref{eq:ISTGeneralExpressionForThePotential}) one may see that the
determinant of the matrix ${\mathbf{I}} + {\mathbf{C}}$ is a
polynomial of the entries, i.e. of exponential functions. Taking the
natural logarithm and differentiating twice with respect to the spatial
variable will not change the fact the potential is a rational
function of a finite number of exponentials. This fact is more apparent
in formula (\ref{eq:FormalPowerSeries}). This gives us hope to be able
to explore analytically the properties of a quantum system which
possess the zeros of the Riemann $\zeta(s)$ function as energy
eigenvalues.

Simplification of formulae (\ref{eq:ISTGeneralExpressionForThePotential}-c)
is under investigation.

\begin{acknowledgments}
   D. S. is grateful to Dr. Dennis McCaughan for the discussion about
   the convergence of a sequence containing the M{\"{o}}bius
   function. This work was financially supported by the New Zealand
   Foundation for Research Science and Technology under the contract
   NERF-UOOX0703: Quantum Technologies. Brandon P. van Zyl acknowledges
   financial support from the Natural Sciences and Engineering Research
   Council (NSERC) of Canada.
\end{acknowledgments}

\end{document}